\documentstyle[aps,epsf,multicol]{revtex}
\title{Turbulence without pressure in $d$ dimensions}
\author{S.~A.~Boldyrev \\
\em{Princeton University, P.O.Box 451, Princeton, NJ 
08543}}
\date{\today}
\begin{document}
\input psfig.sty
\maketitle
\begin{abstract}

The randomly driven Navier-Stokes equation without  
pressure in $d$-dimensional space is considered as a model of strong turbulence 
in a compressible 
fluid. 
We derive a closed equation for the velocity-gradient probability
density function. We find the asymptotics of this function for the case of the
gradient velocity field (Burgers turbulence), and provide a numerical
solution for the two-dimensional case. Application of these results to the 
velocity-difference probability density function is discussed.

~\\
\noindent PACS Number(s): 47.27.Gs, 03.40.Kf, 52.35.Ra.
\end{abstract}

\begin{multicols}{2} 
\section{Introduction}
The Burgers equation with a random external force is considered to be 
the first exactly
solvable model of $1d$~turbulence and has been extensively studied 
in recent years
\cite{Polyakov,Y-Ch,Boldyrev,Boldyrev2,Bouchaud,Sinai,G-M,Bouchaud1,Balkovsky,Gotoh,Ivashkevich}. 
Though rather simplified, this model can serve as a test model for 
some general 
ideas within the theory of strong turbulence. In~1995, methods of 
quantum field 
theory were applied to this problem by A.~Polyakov~\cite{Polyakov} 
which enabled a qualitative explanation of velocity-difference
probability density functions (PDFs) measured numerically 
by A.~Chekhlov and V.~Yakhot~\cite{Y-Ch}. In~\cite{Boldyrev} it was shown 
that the approach~\cite{Polyakov} allows one to obtain  
quantitatively correct 
results. Extensive numerical simulations published recently by T.~Gotoh 
and R.~Kraichnan~\cite{Gotoh} show that the predictions 
of~\cite{Polyakov,Boldyrev} 
are quite accurate, and coincide with the numerical simulations to within 
about~5$\%$. V.~Yakhot has shown in~\cite{Yakhot} that the ideas introduced
in~\cite{Polyakov} can have much wider application, and can also work for
incompressible velocity fluctuations.

We believe that the operator product expansion (OPE), 
introduced in~\cite{Polyakov} to take into account the viscous term, is 
an adequate language to
treat compressible turbulence in higher dimensions as well, where shock
structures and associated local dissipation persist. In the present 
paper we find a closed equation for the velocity-gradient probability-density
function (PDF) for compressible turbulence in any number of dimensions. We 
investigate the
asymptotics of the PDF and present the numerical solution for 
the~$2d$ case.

The basic equation we will study is the following:

\begin{eqnarray}
 {\bf u}_t + ({\bf u}\cdot \nabla){\bf u} = 
\nu \Delta {\bf u} + {\bf f}\,\,. \label{Eq.1}
\end{eqnarray}

The force~$f$ is chosen to be Gaussian with zero mean
and white in time variance:
\begin{eqnarray}
\langle f^i({\bf x}, t)f^k({\bf x}^{\prime}, t^{\prime}) \rangle = \delta(t-t^{\prime})
\kappa^{ik}({\bf x}-{\bf x}^{\prime})\,\,,
\label{Eq.0}
\end{eqnarray}

\noindent where the $\kappa$~function is concentrated at some large 
scale~$L$, and can be expanded as follows  

\begin{eqnarray}
\kappa^{ik}({\bf
y})=\kappa_0\delta^{ik}-\kappa_1\left(y^2\delta^{ik}+2\alpha y^iy^k\right)
 \label{force}
\end{eqnarray} 
\noindent for $y \ll L$. We assume that the steady states 
for velocity gradient and velocity 
difference exist; for this we can require, for example, that periodic boundary 
conditions on a scale much
larger that~$L$ be imposed, and that the zero harmonic in the 
$\kappa$~function be
absent. These assumptions are usually used in numerical simulations \cite{Y-Ch,Gotoh}.

In this paper we appeal to the results obtained for the~$1d$ Burgers 
turbulence
without pressure in~\cite{Polyakov,Y-Ch,Boldyrev,Boldyrev2,Sinai,Gotoh}. In particular, 
we are
interested in the velocity-gradient~PDF $P(\partial u^i/\partial x^k)$ and 
the velocity-difference~PDF $P_v({\bf u}({\bf x}_1)-{\bf u}({\bf
x}_2))$, where the
velocities are taken at the same time at some fixed points~${\bf x}_1$ and~${\bf
x}_2$. The physical picture 
presented in
these papers allows us to consider a general phenomenon such as 
intermittency on a
rigorous basis; it is related to the spontaneous breakdown of the 
Galilean 
invariance of the forced equation and to the algebraic decay of the 
PDFs. We will not
repeat these arguments here; instead, we will concentrate on the 
main ideas
which allow us to consider the multi-dimensional case. 

We will be interested in the
case of small dissipation $\nu$, and will consider distances $|x_1-x_2|\ll L$. The
following order of the limits should be considered to get the steady state: we first
set $t\rightarrow \infty$, and then consider the limit $\nu \rightarrow 0$.

\section{Velocity-gradient PDF}

To proceed quantitatively, consider the following characteristic
function ($Z$ function)  for the
velocity gradient $u^i_k\equiv \partial u^i/\partial x^k$:

\begin{eqnarray}
Z(\sigma_{kl})\equiv \langle \exp(i\sigma_{kl}u^k_l) \rangle \,\,\,. \label{Zfunction}
\end{eqnarray}

Due to Eq.~(\ref{Eq.1}), this function satisfies the following Fokker-Planck
equation:

\begin{eqnarray}
& {\dot Z} =i\sigma_{ik}\frac{\partial^2 Z}{\partial \sigma_{lk}\partial\sigma_{il}}
- i\delta_{ik}\frac{\partial Z}{\partial
\sigma_{ik}}- \nonumber \\
&
-\left[\alpha
\sigma_{ii}\sigma_{kk}+\frac{1+\alpha}{2}\sigma_{ik}\left(\sigma_{ik}+\sigma_{ki}
\right)+\right.\nonumber \\
& \left. \frac{1-\alpha}{2}\sigma_{ik}\left(\sigma_{ik}-\sigma_{ki}
\right) \right]Z+D,
\label{FP}
\end{eqnarray}
\noindent where summation over repeated indices is assumed. For 
simplicity, we set $\kappa_1=1$. To derive this equation we
differentiated~(\ref{Zfunction}) with respect to~$t$ and expressed~$\dot u^i_k$ 
using
Eq.~(\ref{Eq.1}). We made use of the following identities:
\begin{eqnarray}
\langle i\sigma_{ik} u^l_k u^i_l\exp(i\sigma_{mn}u^m_n)\rangle = 
- i\sigma_{ik}\frac{\partial^2}{\partial \sigma_{lk}\partial\sigma_{il}}Z \,\,,
\end{eqnarray}
 
\begin{eqnarray}
 & \langle i\sigma_{ik} u^l u^i_{lk}\exp(i\sigma_{mn}u^m_n)\rangle = \langle u^l
\frac{\partial}{\partial x^l}\exp(i\sigma_{mn}u^m_n) \rangle = \nonumber \\ 
 & - \langle u^l_l \exp(i\sigma_{mn}u^m_n) \rangle = 
i\delta_{ik}\frac{\partial}{\partial \sigma_{ik}}Z\,\,,
\end{eqnarray}
\noindent and 
\begin{eqnarray}
\langle i \sigma_{ik} f^i_k \exp(i\sigma_{mn}u^m_n) \rangle = 
\frac{1}{2}\, \sigma_{ik}\sigma_{mn}\kappa^{im}_{kn}(0)\, Z\,\,.
\end{eqnarray}

\noindent The~$D$ term 
describes the contribution of the dissipation and in steady 
state is given by: 

\begin{eqnarray}
D=\lim \limits_{\nu\rightarrow 0} \langle \nu i \sigma_{rs}\Delta u^r_s 
\exp(i\sigma_{kl}u^k_l)\rangle\,\,\,. \label{Dterm}
\end{eqnarray}

Without this term the steady state in Eq.~(\ref{FP}) does not exist. This term can
not be closed without further assumptions. In \cite{Polyakov}, assumptions about 
scaling invariance, Galilean invariance, and the operator product expansion  
were applied 
to close the analogous term for the velocity-{\em difference} PDF. It is not obvious 
{\em a priori} that these methods can be applied to our problem, since the 
limits~$y\rightarrow 0$ and~$\nu \rightarrow 0$ may not be interchangeable. 

Nevertheless, it was observed in the
numerical simulations in~\cite{Gotoh} that the
velocity-difference and the velocity-gradient PDFs coincide 
for the Galilean invariant region~$\Delta u \ll u_{rms}$
($u_{rms}=(\kappa_0L)^{1/3}$) in the 
one-dimensional case. 
This
suggests that the velocity-difference PDF is contributed to by {\em smooth} parts of 
the velocity field in the Galilean invariant region, and therefore the 
limits~$y\rightarrow 0$ and~$\nu \rightarrow 0$ {\em are} interchangeable. 

Another important result of \cite{Gotoh} is that the $\beta$~anomaly 
introduced in~\cite{Polyakov}
is absent for regular forcing.
We assume that this is true for the multi-dimensional case as well. Under this
assumption, the~$D$ term in the multi-dimensional case should be expanded as 
\begin{eqnarray}
D=aZ\,\,.\label{OPE}   
\end{eqnarray}

This is the only assumption we use in what follows. We refer the reader to
papers~\cite{Polyakov,Y-Ch,Boldyrev,Gotoh} for more details and discussions on 
the underlying ideas. 
We will see that this assumption is self-consistent; the anomaly $a$ can be 
found from the conditions of positivity, finiteness, and normalizability of the PDF. 
These 
conditions
can be easily imposed on the~PDF in $u$-space. We therefore transform Eq. (\ref{FP}) to
 $u$-space, using: 
\begin{eqnarray}
P(\nabla u)=\int \mbox{d}\sigma  Z(\sigma)e^{-i\sigma_{kl}u^k_l}
\,\,,\label{Fourier}
\end{eqnarray}
\noindent where $\mbox{d}\sigma =\prod_{i,k}\mbox{d}\sigma_{ik}$ is the 
 measure in  
$d^2$-dimensional space of the elements of the matrix~$\sigma_{ik}$. 
In the steady state, the equation takes the form:
\begin{eqnarray}
 & u^i_iP + \frac{\partial }{\partial u^i_l}\left(u^i_ku^k_l P\right)+ \nonumber \\
 & +\left[
\alpha \frac{\partial^2}{\partial u^i_i\partial u^k_k} + 
\frac{1+\alpha}{2}\frac{\partial}{\partial u^i_k}\left(\frac{\partial}{\partial u^i_k}+
\frac{\partial}{\partial u^k_i}\right)+\right.\nonumber \\
& \left. +
\frac{1-\alpha}{2}\frac{\partial}{\partial u^i_k}\left(\frac{\partial}{\partial u^i_k}-
\frac{\partial}{\partial u^k_i}\right)
\right]P=-aP. \label{FP1}
\end{eqnarray}
This is the general equation for the~PDF. The force in this equation
can have different symmetry properties, which correspond to different
values of the parameter~$\alpha$ in~(\ref{force}).  No restrictions have so far 
been imposed on the velocity field either.

Eq.~(\ref{FP1}) can be simplified 
for
the {\em gradient} force~${\bf f}=\nabla \phi$. This choice 
corresponds to~$\alpha=1$ and 
allows us to look for a solution in the factorized form:
\begin{eqnarray}
P=\left[\prod\limits_{i<k}\delta(u^i_k-u^k_i)\right]\,{\tilde P}({\tilde u})\,\,,
\label{factorization}
\end{eqnarray}
\noindent where~${\tilde u}$ denotes the symmetric part of the matrix~$u_{ik}$.
Physically, this means that we have restricted our consideration to  {\em gradient}
fluctuations of the velocity field, ${\bf u}=\nabla h$. We refer to this case as
 multi-dimensional Burgers turbulence. It was considered by completely
different methods in~\cite{Bouchaud,G-M}. The spirit of our method is most close to
the consideration of~\cite{Bouchaud1}. 
Eq.~(\ref{FP1}) with ansatz~(\ref{factorization})
can be cast into the following form: 
\begin{eqnarray}
 & (d+2){\tilde u}^i_i{\tilde P} + {\tilde u}^i_k{\tilde u}^k_l\frac{\partial}
{\partial u^i_l} {\tilde P}+ \nonumber \\
 & +\left[
\frac{\partial^2}{\partial u^i_i\partial u^k_k} + 
2\frac{\partial^2}{\partial u^i_k \partial u^i_k}
\right]{\tilde P}=-a{\tilde P}. \label{FP2}
\end{eqnarray}
\noindent In what follows we will consider only the function~${\tilde P}$, and 
will omit the overtilde sign.

Eqs.~(\ref{FP1}) and~(\ref{FP2}) help to reveal the physical sense of the $a$~anomaly.
Integrating these equations with respect to~$u^i_k$ one gets:

\begin{eqnarray}
\langle u^i_i \rangle = -a,
\end{eqnarray}

\noindent which means that this anomaly describes the average measure loss due to
compressibility and presence of shocks. It can also be interpreted as the mean rate of
density accumulation on shocks in the Lagrangian picture:

\begin{eqnarray}
& {\dot u}^i_k(y,t) + u^i_k u^k_i = f^i_k, \nonumber \\
& {\dot \rho} (y,t)+ u^i_i \, \rho (y,t) + a \, \rho (y,t) = 0, \label{Langevin}
\end{eqnarray}

\noindent where $y$ is the Lagrangian coordinate,  $u^i_k$ represents the smooth part 
of the velocity field, and $a=\langle u^i_i\rangle_{shocks}$. The  Lagrangian
PDF~$P(u^i_k, \rho \,; y)$ and  the Eulerian PDF~$P(u^i_k; x)$ are related as follows:

\begin{eqnarray}
P(u^i_k; x)=\int \frac{\mbox{d}\rho}{\rho} \, P(u^i_k, \rho \, ; y), 
\label{Euler-Lagrange}
\end{eqnarray}

\noindent that leads to Eq.~(\ref{FP1}). We note an 
interesting analogy with a mean field approximation: $a$ is introduced as the mean 
field in the dynamical equations and then is found self-consistently from
Eq.~(\ref{FP2}). The interpretation~(\ref{Langevin}) is important, since it
allows one to introduce the anomaly on the level of the stochastic Langevin equation.

In general, the PDF should depend only on invariants with respect to 
space rotations. For the $d$-dimensional space, there are exactly~$d$ such 
invariants, which 
can be chosen as the eigenvalues of the matrix~${\tilde u}_{ik}$. Let us denote them 
as $\lambda_1,\lambda_2, \dots , \lambda_d$. Eq. (\ref{FP2}) 
can be rewritten for the function $P$ depending on only these variables:

\begin{eqnarray}
& \sum\limits_{k=1}^{d} \left[(d+2)\lambda_k +\lambda_k^2 \frac{\partial}
{\partial \lambda_k}\right]P + \left( \sum\limits_{k=1}^{d}\frac{\partial }
{\partial \lambda_k} \right)^2 P \nonumber \\
& + 2 \sum\limits_{k=1}^{d} \left( \frac{\partial}
{\partial \lambda_k}\right)^2P 
+\sum\limits_{i,k}\frac{1}{\lambda_i-\lambda_k}\left(\frac{\partial}{\partial
\lambda_i}-\frac{\partial }{\partial \lambda_k}\right)P=-aP.
\label{FP3} 
\end{eqnarray}
\noindent To derive this equation we used the following expression 
for the matrix
Laplacian, known in the theory of matrix models~\cite{Mehta,Newman}:
\begin{eqnarray}
\nabla^2_{\tilde u}=\sum\limits_i\frac{\partial^2}{\partial \lambda^2_i} +
\frac{1}{2}\sum\limits_{i,k}\frac{1}{\lambda_i-\lambda_k}
\left( \frac{\partial}{\partial
\lambda_i}-\frac{\partial }{\partial \lambda_k}\right)\,\,.\label{Laplacian}
\end{eqnarray}

Equation~(\ref{FP3}) has an infinite number of solutions. The physically reasonable
solution should satisfy conditions of positivity, finiteness, and normalizability;
exactly in the same manner as the ground state is determined in quantum mechanics.
The solution should also be symmetrical with respect to the 
arguments~$\lambda_1,\dots,\lambda_d$. These conditions should determine the 
unknown parameter~$a$. This parameter depends only on the symmetry properties 
of the external force and on the space dimensionality.

As in the one-dimensional
case, the asymptotics of the solutions can be found by balancing different 
terms in eq. (\ref{FP3}).
If we balance the advective and force terms, we will get the PDF tail in the region
where the dissipative effects are negligible. In the one-dimensional case this
corresponds to the right tail of the  PDF. This tail decays hyper-exponentially fast. 
In the multi-dimensional case the analogous asymptotic should have the form:
$P\propto \exp(S(\lambda_1,\dots, \lambda_d))$. The function~$S$ should be 
symmetric with respect to its arguments~$\lambda_1,\dots,\lambda_d$. The asymptotic
can be simply found for large positive $\lambda$'s in the direction close 
to $\lambda_1=\dots=\lambda_d$:
\begin{eqnarray}
P \propto \exp\left[\frac{-\Lambda^3}{3d^2(d+2)} \right]\,\,, 
\label{asymptotics1}
\end{eqnarray}
\noindent where $\Lambda \equiv \mbox{Tr}({\tilde u}_{ik})=\lambda_1+\dots+\lambda_d$. 
The same asymptotic for large $\lambda$ can also be obtained by the instanton
methods~\cite{G-M,Bouchaud1} applied directly to quantum mechanics~(\ref{FP3}).

The tail, corresponding to  large negative~$\lambda$'s (the ``left" tail)  decays
rather slowly. The explanation is simple. Burgers
shocks always have negative velocity jumps, and therefore large positive velocity
gradients are less probable than large negative ones. The left tail is determined
by large negative gradients, and to obtain it we should neglect the force term 
in~(\ref{FP3}). We find:
\begin{eqnarray}
\label{left}
P\propto
\frac{G(I_{ik})}{(\lambda_1\dots\lambda_d)^{d+2}}\,\,,
\end{eqnarray}
\noindent where $G$~is some function, and 
$I_{ik}=(\lambda_i-\lambda_k)/\lambda_i\lambda_k$ are
$d(d-1)/2$~invariants of the characterictic equations for~(\ref{FP3}). The
finite solution which is non-vanishing for $\lambda_1=\dots=\lambda_d$ has the form:

\begin{eqnarray}
P\propto
\frac{1}{(\lambda_1\dots\lambda_d)^{d+2}}\equiv
\mbox{Det}^{-(d+2)}({\tilde u}_{ik})\,\,.\label{asymptotics2}
\end{eqnarray}

The function obtained from (\ref{FP3}) should be normalized with respect to the 
flat measure in the $d(d+1)/2$-dimensional space of elements of the
symmetric matrix~${\tilde u}_{ik}$. In 
$\lambda$-space, this normalization is performed as follows:

\begin{eqnarray}
\int P(\lambda)\, |\Delta(\lambda)|
\prod\limits_{k=1}^d \mbox{d}\lambda_k =1,
\label{normalization}
\end{eqnarray}

\noindent where $\Delta(\lambda)=\prod_{i<j}\left(\lambda_i-\lambda_j\right)$ is the
VanderMonde determinant; for 
details, see \cite{Mehta,Newman,Z-J}.

\section{Numerical solution for the two-dimensional case}

In this section we solve Eq. (\ref{FP3}) numerically in the two-dimensional case. The
purpose of these calculations is to show that Eq.~(\ref{FP}) with the anomaly
term~(\ref{OPE}) does have a 
steady state, at least for the gradient velocity field.

We have used the
relaxation method and started with some arbitrary but {\em symmetrical} initial
distribution. The numerical
value for the anomaly turned out to be~$a= 1.30\pm0.02$. The PDF has 
hyper-exponential and power-like tails and is presented in~Fig.~\ref{fig1}. The PDF is
normalized according to (\ref{normalization}). Plotted on the  horizontal axes 
are~$\lambda_1$~and~$\lambda_2$.
{\columnwidth=3in
\begin {figure} [tbp]
\centerline {\psfig{file=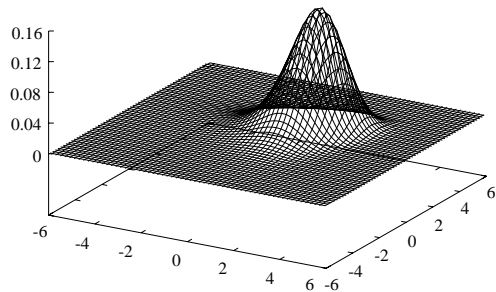,width=8cm}}
\caption{ Velocity-gradient probability-density function $P(\lambda_1, \lambda_2)$.}
\label{fig1}
\end{figure}
}
Fig.~\ref{fig2} shows the
same PDF for the diagonal direction $\lambda_1=\lambda_2$. The left tail 
decays as~$1/\Lambda^8$, the
right tail asymptotic is~$P\propto \exp(-\Lambda^3/48)$, in agreement
with~(\ref{asymptotics1}) 
and ~(\ref{asymptotics2}).  $P(\Lambda)$~is plotted vs.
$\sqrt{\lambda_1^2+\lambda_2^2}=\Lambda/\sqrt{2}$.
{\columnwidth=3in
\begin {figure} [tbp]
\centerline {\psfig{file=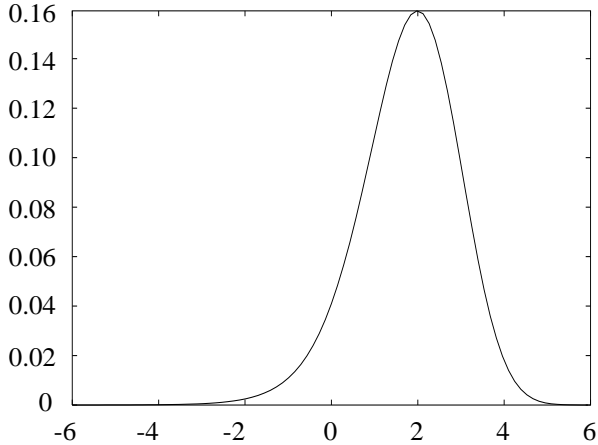,width=8cm}}
\vskip3mm
\caption{Velocity-gradient~PDF $P(\Lambda)$ for the diagonal direction 
$\lambda_1=\lambda_2$.}
\label{fig2}
\end{figure}
}
\section{Concluding remarks}

The crucial assumption in our treatment of the dissipative anomaly is the assumption
that only smooth parts of the velocity field contribute to the
velocity-{\em difference}~PDF. After the velocity-gradient 
$Z$~function~(\ref{Zfunction}) is
found, the velocity-difference $Z$~function can be constructed as follows:

\begin{eqnarray}
Z_v(\zeta_i, y^k)=Z(\zeta_i y^k)\equiv\langle \exp(i\zeta_i y^ku^i_k) 
\rangle \,\,\,, \label{Zvfunction}
\end{eqnarray}

\noindent i.e., we simply changed $\sigma_{ik}\rightarrow\zeta_i y_k$
in~(\ref{Zfunction}). 
The Fourier transform with respect to $\zeta$ will then give the
velocity-difference~PDF. As an example, let us consider the longitudinal
velocity-difference PDF:

\begin{eqnarray}
& P_l(\Delta u, y)\propto \nonumber \\
&  \int \mbox{d}u_{ik}\mbox{d}\zeta P(\lambda)\exp
\{ i\zeta y n_i n_k u_{ik}-i\zeta \Delta u \},
\label{P_l-definition}
\end{eqnarray}
\noindent where~$n_i$ is a unit vector in the direction of~$y_i$. Since~$P_l(\Delta
u, y)$ does not depend on~$n_i$, one can average with respect to all possible
directions of this vector and get the following result:

\begin{eqnarray}
\label{P_l-integral}
& P_l(\Delta u, y)\propto \nonumber \\
& \frac{1}{y}\int P(\lambda)|\Delta(\lambda)|
\delta \left( \lambda_i n_i^2 - \frac{\Delta u}{y} \right )\delta (1-{\bf
n}^2)\prod\limits^d_{i=1} \mbox{d}\lambda_i \mbox{d} n_i .
\end{eqnarray}

In the two-dimensional case this integral can be simplified further:

\begin{eqnarray}
P_l(\Delta u, y) \propto \frac{1}{y} \int\limits^{\Delta u/y}_{-\infty }
\int\limits^{\infty }_{\Delta u/y} \frac{\mbox{d} \lambda_1 \mbox{d} \lambda_2 
\vert \lambda_1-\lambda_2 \vert P(\lambda)}
{\left[ \left( \frac{\Delta u}{y}-\lambda_1 \right)\left(\lambda_2 - 
\frac{\Delta u}{y} \right)\right]^{1/2} }
\end{eqnarray}

\noindent In general,~$P_l(\Delta u, y)$ can be represented 
as~$P_l(\Delta u, y)=w(\Delta u/y)/y$.  If we assume that for large 
negative~$\Delta u/y $ the integral~(\ref{P_l-integral}) is
contributed to by the force-free asymptotic~(\ref{left}), we immediately 
get the left tail asymptotic for the longitudinal velocity-difference PDF:  

\begin{eqnarray}
w(z)\propto z^{-(d+1)(d+2)/2}, \,\,\, z \to -\infty.
\end{eqnarray}
Analogously, one can obtain a PDF for $\nabla \cdot u$. For this
purpose one should set $\sigma_{ik}\rightarrow\delta_{ik}\zeta$. Such 
a PDF was investigated numerically in~\cite{Gotoh1}, though the~$Re$ 
number was not large enough to obtain the inertial range.

Finally, we would 
like to note that the absence of the $\beta$~anomaly, that we assumed in
our consideration, can be not a
universal fact. It was conjectured in~\cite{Boldyrev} that different 
dissipative regularizations (e.g. hyper-dissipation
$(-1)^p\partial^{2p}/\partial x^{2p}$) can lead to
different steady states. This assumption is natural for the language of the 
OPE: different
dissipative operators should have different expansion coefficients~$a$ and~$b$ (we
use the notation of \cite{Polyakov}). Moreover, some analog of the $\beta$ 
anomaly can also be present in Eq.~(\ref{FP1}), since it describes a general 
velocity field, without ``gradient" restriction~(\ref{factorization}).

These questions are under consideration, the results will be reported
elsewhere.

\vskip5mm

I am indebted to A.~Polyakov and V.~Yakhot for many important 
discussions and comments. 
I would also like to thank T.~Gotoh, V.~Gurarie, and R.~Kraichnan for useful 
conversations, D.~Uzdensky for helpful discussions on both the physics and the 
numerics of the problem, and T.~Munsat for valuable remarks on the style 
of the paper.

This work was supported by U.S.D.o.E. Contract 
No. DE--AC02--76--CHO--3073.

\end{multicols}


\vskip5mm

\begin{thebibliography}{99}


\bibitem{Polyakov} A.~Polyakov, Phys. Rev. E {\bf 52}, 6183 (1995); 
hep-th/9506189.

\bibitem{Y-Ch} A.~Chekhlov and V.~Yakhot, Phys. Rev. E {\bf 52}, 
5681 (1995); V.~Yakhot and A.~Chekhlov, Phys. Rev. Lett. {\bf 77}, 
3118 (1996).


\bibitem{Boldyrev} S.~Boldyrev, Phys. Rev. E {\bf 55}, 6907 (1997); 
hep-th/9610080.
 


\bibitem{Boldyrev2} S.~Boldyrev, Phys. Plasmas {\bf 5}, 1681 (1998).


\bibitem{Bouchaud} J.-P.~Bouchaud, M.~M\'ezard, and G.~Parisi, 
Phys. Rev. E {\bf 52}, 3656 (1995).

\bibitem{Sinai} W.~E, K.~Khanin, A.~Mazel, and Ya.~Sinai, Phys. 
Rev. Lett. {\bf 78}, 1904 (1997).

\bibitem{G-M} V.~Gurarie and A.~Migdal, Phys. Rev. E {\bf 54}, 
4908 (1996); hep-th/9512128.

\bibitem{Bouchaud1} J.-P.~Bouchaud, M.~M\'ezard, {\em Velocity Fluctuations in Forced
Burgers Turbulence}, cond-mat/9607006, July 1996.

\bibitem{Balkovsky} E.~Balkovsky, G.~Falkovich, I.~Kolokolov, and 
V.~Lebedev, Phys. Rev. Lett. {\bf 78}, 1452 (1997); chao-dyn/9609005.

\bibitem{Gotoh} T.~Gotoh and R.~H.~Kraichnan, Phys. Fluids {\bf 10} (1998) 2859; 
chao-dyn/9803037.

\bibitem{Ivashkevich} E.~Ivashkevich, J. Phys. A {\bf 30} (1997) L525; 
hep-th/9610221.

\bibitem{Yakhot} V.~Yakhot, Phys. Rev. E {\bf 57} (1998) 1737; chao-dyn/9708016.

\bibitem{Mehta} M.~L.~Mehta, {\em Random Matrices} (Academic Press, Boston, 1991),
2nd~ed.

\bibitem{Newman} M.~Newman, {\em Matrix Models Coupled to an External Field},
Ph.~D.~Thesis, Princeton University, 1992.

\bibitem{Z-J} J.~Zinn-Justin, {\em Quantum Field Theory and Critical Phenomena},
Clarendon Press, Oxford, chapter A28, 1993.

\bibitem{Gotoh1} T.~Gotoh, Forma {\bf 8}, 133 (1993).


\end{thebibliography}
\end{document}